\documentclass[12pt,titlepage]{article}
\usepackage{amssymb}
\usepackage{epsfig}
\usepackage{bm}
\font\grande=cmr10 scaled \magstep4
\font\medio=cmr10 scaled \magstep2
\outer\def\beginsection#1\par{\medbreak\bigskip
      \message{#1}\leftline{\bf#1}\nobreak\medskip
\vskip-\parskip
      \noindent}

\newcommand{\eq}{\begin{equation}}

\newcommand{\eqx}{\end{equation}}
\newcommand{\eqn}{\begin{eqnarray}}

\newcommand{\bi}{\begin{itemize}}

\newcommand{\eqnx}{\end{eqnarray}}
\newcommand{\ei}{\end{itemize}}

\newcommand{\nn}{\nonumber}

\newcommand{\kt}{\boldsymbol{k}}

\newcommand{\xt}{\boldsymbol{x}}

\newcommand{\A}{\mathcal{A}}
\newcommand{\Hc}{\mathcal{H}}

\begin{document}
\begin{center}

\grande{Exploring an S-matrix for gravitational collapse \\
II: a momentum space analysis}

\vskip 5mm
\vskip 5mm
\large{ G. Veneziano}
\vspace{3mm}

{\sl Theory Division, CERN, CH-1211 Geneva 23, Switzerland}

{\sl and}

{\sl Coll\`ege de France, 11 place M. Berthelot, 75005 Paris, France}

\vspace{6mm}

\large{ J. Wosiek}
\vspace{3mm}

   {\sl M. Smoluchowski Institute of Physics, Jagellonian University}

{\sl Reymonta 4, 30-059 Cracow, Poland}\\

 \vspace{6mm}
\centerline{\medio  Abstract}
\vskip 5mm
\end{center}
 We complement our earlier position-space exploration of a recently proposed
 S-matrix for transplanckian scattering by a momentum-space analysis.
 As in the previous paper,  we restrict ourselves to the   case of axisymmetric
 collisions of extended sources.  Comparison between the two formulations allows
 for several cross-checks while showing their complementary advantages.
 In particular, the momentum-space formulation leads to an easier computation
 of the emitted-graviton spectra and to an attempt to study the system
 beyond its critical points into the presumed  gravitational-collapse
 regime.

 \vspace{5mm}

\vfill
\begin{flushleft}
CERN-PH-TH/2008-109 \\
TPJU-4/2008 \\
May  2008\\
\end{flushleft}
\vfill
\section{Introduction}
In a companion paper \cite{VW08} we have explored a recent
proposal \cite{ACV07} for  an $S$-matrix description of
transplanckian scattering in four-dimensional spacetime. The
proposal is based on an approximate resummation of the
semi-classical corrections to the leading eikonal approximation
\cite{ACV} and
 amounts to solving the equations of motion of an effective action introduced quite sometime ago \cite{Lip, ACV93}. In a suitable limit,  the longitudinal dynamics  can be factored out leaving behind an effective dynamics in the ``transverse" $two$-dimensional space.

 In \cite{VW08}  we considered, quite systematically, the case of the axisymmetric (central) collision of two extended sources/beams. This case has several advantaged over the original problem of  a  two-particle collision at non-vanishing impact parameter:
 \begin{itemize}
 \item The partial differential equations (PDE's) reduce to  ordinary differential
 equations (ODE's) making the problem affordable by analytic and numerical
 techniques, without having to make either an azimuthal-average
 approximation \cite{ACV07} or to resort to advanced numerical techniques \cite{MO} .
 \item The IR-sensitive graviton polarization, which was neglected by hand in \cite{ACV07},  is simply not produced in the axisymmetic case \cite{VW08}.
 \item  We  can consider a wide variety  of initial states
 by playing with the many (shape and intensity) parameters chracterizing
 the sources and check for the existence of critical surfaces in this
 multidimensional space. The results can then be compared with those coming
 from closed trapped surface (CTS) criteria \cite{EG}, \cite{KV} and will be tested,
  hopefully in the near future,  against  numerical GR calculations
  (see  \cite{AE}, \cite{numerical} for a few results already available for this case).
 \end{itemize}
The results of \cite{VW08}, based on a position-space analysis, gave further support to the conclusions of \cite{ACV07}.
In particular, we were able to prove a one-way relation between the CTS criterion of \cite{KV} and the criticality condition in the ODE system. We could also determine quite precisely the critical surfaces  in a variety of cases and found a good quantitative agreement with CTS-based criteria  \cite{KV}.  Finally, we confirmed that, above those critical lines,  a new absorption of the elastic S-matrix turns on with a universal  behaviour  reminiscent of Choptuik's critical exponent \cite{Chop}.

In this paper we complement our previous work  \cite{VW08} by a momentum-space analysis according to the following outline:   in section 2 we present the momentum space formulation of the extended-source problem  and, after   specializing to the axisymmetric case, we give the explicit form of the action and of the equations of motion.  In section 3 we recall some interesting extended sources already considered in \cite{VW08}  adding their  momentum-space form. In  section 4 we study  numerically the field equations, determine the critical lines in  parameter space, and compare them with the position-space results of \cite{VW08}.  In section 5 we discuss the spectrum of the emitted gravitons starting from the perturbative regime and until one approaches the critical lines.
In section 6  we present  an attempt to extend the solutions  in the (presumed) BH-phase.

\section{Momentum space action and field equations}

We recall from \cite{VW08} the position-space action of \cite{ACV07} generalized to extended sources:
\eqn
\label{Aextx}
\frac{\A}{2 \pi G s} &=& \int d^2x \left[a(x) {\bar s}(x) + {\bar a}(x) s(x) - \frac12 \nabla_i {\bar a} \nabla_i a \right] \nonumber \\
&-& \frac{(\pi R)^2}{2}  \int d^2x \left((\nabla^2 \phi)^2
+2 \phi (\nabla^2 a~\nabla^2\bar{a}-\nabla_i\nabla_j a~\nabla_i\nabla_j\bar{a}) \right)\, ,
\eqnx
with the three real fields $a$, $\bar{a}$ and $\phi$ representing
the two longitudinal and the single, IR-safe, transverse component of the gravitational
field, respectively.

The center of mass energy ${\sqrt s}$ provides the overall normalization factor $ 2 \pi G s = \frac{\pi}{2 G} R^2$, while the two sources $s(x), {\bar s} (x)$ are normalized by $\int d^2x ~s(x)= \int d^2x~ {\bar s} (x)  = 1$.
In order to go to momentum space we can either start from (\ref{Aextx}) or generalize directly  eq. (5.2) of \cite{ACV07}. The result is:
\eqn
\label{pspA}
\frac{\pi A}{Gs} &=&  \int \frac{d^2\kt}{\kt^2} \left[  \beta_1(\kt)s_2(-\kt) +  \beta_2(\kt)s_1(-\kt) -  \beta_1(\kt) \beta_2(-\kt)  \right]  \nn \\
 &-& \frac{(\pi R)^2}{2} \int d^2 \kt  \left[  \frac{1}{2} h(\kt)  h(-\kt) -  h(-\kt) \Hc(\kt)  \right] \, ,
\eqnx
where the FT of the sources (still denoted by $s_i$)  are normalized by requiring $s_i(0) =1$.  Furthermore,
\begin{equation}
\beta_1(\kt) = \frac{k^2 a(\kt)}{2}~~~,~~~ \beta_2(\kt) = \frac{k^2 {\bar a}(\kt)}{2}~~,~~
h(\kt) = -  k^2  \phi(\kt) \, ,
\end{equation}
\begin{equation}
 \Hc(\kt) \equiv  \frac{1}{\pi^2 \kt^2} \int d^2\kt_1 d^2\kt_2 \delta(\kt-\kt_1-\kt_2) \beta_1(\kt_1)  \beta_2(\kt_2)   sin^2 \theta_{12} \, ,
 \end{equation}
and $\theta_{12}$ is the angle between the two transverse momenta $\kt_1$ and $\kt_2$.

If we now specialize to the axisymmetric case where sources and fields depend only upon $k_i^2$,  we can use the following relations:
\eqn
\int d^2k &=& \pi \int d k^2~~,~~ \nonumber \\   \int d^2\kt_1 d^2\kt_2 \delta(\kt-\kt_1-\kt_2) \frac{sin^2 \theta_{12}}{k^2} &=&
\frac14 \int_{\lambda \ge 0} \frac{dk_1^2 dk_2^2}{k_1^2 k_2^2 k^2} \sqrt{\lambda(k_1^2, k_2^2, k^2)} \, ,
\eqnx
where
\eq
\lambda(k_1^2, k_2^2, k^2) = (2 k_1^2 k_2^2 +2 k^2 k_2^2+  2 k_1^2 k^2 - k^4 -k_1^4 - k_2^4)\, ,
\eqx
and we have used the
 identity:
\eq
sin^2 \theta_{12} = \frac{\lambda(k_1^2, k_2^2, k^2)}{4 k_1^2 k_2^2} \, ,
\eqx
together with the change of variable (at fixed $k^2$, $k_1^2$)
\eq
d k_2^2 = d (k-k_1)^2 = 2 |k||k_1| \sin \theta_{k1} d \theta_{k1} = 2 |k|_1|k_2| \sin \theta_{12} d \theta_{k1} \, .
\eqx
One thus arrives at the final form of the
 axially-symmetric effective action:
\eqn
\label{pspA}
&& \frac{A}{Gs} =  \int \frac{dk^2}{k^2} \left[  \beta_1(k^2)s_2(k^2) +  \beta_2(k^2)s_1(k^2) -  \beta_1(k^2) \beta_2(k^2)  \right]  \nn \\
 &-& \frac{(\pi R)^2}{2} \int dk^2  \frac{1}{2} h(k^2)  h(k^2) + \frac{R^2}{8}   \int \frac{dk^2 dk_1^2 dk_2^2}{k_1^2 k_2^2 k^2} \sqrt{\lambda(k_1^2, k_2^2, k^2)} h(k^2) \beta_1(k_1^2) \beta_2(k_2^2) \, ,
\eqnx
whose equations of motion read:
\eqn
\label{mseqs}
h(k^2) &=&  \frac{1}{4 \pi^2}  \int \frac{dk_1^2 dk_2^2}{k^2 k_1^2 k_2^2 } \sqrt{\lambda(k_1^2, k_2^2, k^2)}  \beta_1(k_1^2) \beta_2(k_2^2) \, , \nonumber \\
\beta_i(k^2) &=& s_i(k^2) +  \frac{R^2}{8}   \int \frac{dk_1^2 dk_2^2}{k_1^2 k_2^2} \sqrt{\lambda(k_1^2, k_2^2, k^2)} h(k_1^2) \beta_i(k_2^2)\, .
\eqnx
 It is not a completely trivial exercise to show directly the
 equivalence of these equations with the corresponding ones in
 position-space  \cite{VW08} (where a dot stands for  $d / d r^2 $),
  \eqn
 \label{aseqs}
 \dot{a}_i &=& - \frac{1}{2 \pi \rho(r)}\frac{R_i(r)}{R} \, ,\nonumber \\
 \ddot{\rho} &=& \frac12 (2 \pi R)^2 \dot{a}_1 \dot{a}_2 = \frac12 \frac {R_1(r)R_2(r)}{\rho^2(r)}, \nonumber \\
 R_i(r) &=&  R  \int_{|{\bf x}|^2 \le r^2} d^2x ~ s_i(x)\, .
 \eqnx
  The proof, not reported here, makes use of the following (known?) integral of three  Bessel functions (that we have checked numerically):
 \eq
\label{3Jint}
\int_0^{\infty} dr J_1(rk) J_1(rk_1) J_1(rk_2)  =  \frac{1}{2 \pi} \frac{ \sqrt{\lambda(k, k_1, k_2)}}{k k_1 k_2} \Theta(\lambda) \, .
\eqx

\section{Examples of  source profiles}
In this section we list various extended sources already introduced
in \cite{VW08} and give their momentum representations.

{\bf A.}  As a first class, consider finite-size sources
with  the following profiles:
\eq\label{acv}
s_1({\bf x}) = s_2({\bf x}) = \frac{L^4 d}{\pi \left( L^4 d + r^4 (1-d)\right)^{3/2}}\Theta(L-r)\, .
\eqx
Later, without lack of generality, we shall be fixing the transverse size of the
two identical beams $L$ to be 1.
One can easily verify that these sources satisfy our normalizations and that
\eq
\pi \int_0^{r^2} d {\rho}^2 s(\rho) = R(r/L)/R =  \frac{r^2}{ \left( L^4 d + r^4 (1-d)\right)^{1/2}}   ~~, ~~ \pi \int_0^{L^2} s(r)d r^2 = 1\, .
\eqx

Once Fourier transformed to momentum space and
normalized according to the prescription of section 2, the above sources become ($k=|{\bf k}|$):
\eq
s_1(k) = s_2(k) = \frac{\int_0^{L^2} dr^2 J_0(kr) s(r)}{\int_0^{L^2} dr^2  s(r)} = \int_0^{L^2} dr^2 J_0(kr) \frac{L^4 d }{ \left(L^4 d + r^4(1-d)\right)^{3/2}} \, .
\eqx
In particular, for two homogeneous beams ($d=1$) we have:
\eq
s_1(k) = s_2(k) = \int_0^{L^2} dr^2 J_0(kr) = \frac{2}{k L} J_1(k L) \, .
\eqx

{\bf B.} Point-like sources are difficult to deal with numerically,
especially in  momentum space. We introduce therefore
 Gaussian-smeared versions of the  point and
ring-like  sources considered in \cite{VW08}:
\eqn
s_1(\xt)=\frac{1}{{\cal N}_1} \exp{\left(-\frac{r^2}{2\sigma^2}\right)}\Theta&(&L_1-r) ~,~ s_2(\xt)=\frac{1}{{\cal N}_2}
\exp{\left(-\frac{(r-L_2)^2}{2\sigma^2}\right)}\Theta(L_2-r),\;\;\; \nonumber \\
{\cal N}_1=2\pi\sigma^2(1-\exp{\left(\frac{-L_1^2}{2\sigma^2}\right)}) &,&
{\cal N}_2=2\pi\left(\sigma^2(\exp{\left(\frac{-L_2^2}{2\sigma^2}\right)}-1)+
\sigma L_2 \sqrt{\frac{\pi}{2}}   \rm{Erf}{\frac{L_2}{\sqrt{2}\sigma}}
\right)
\eqnx
When $\sigma\rightarrow 0 ~ (\infty)$ such configuration reduces to the one of the point-ring (two homogeneous beams) case.
The corresponding  Fourier transforms are:
\eqn
s_i(\kt)=2\pi \int_0^{L_i} r dr J_0(k r) s_i(\xt)\, .
\eqnx

{\bf C.} Another interesting example is that of gaussian sources concentrated at $r=0$. They correspond to:
\eq \label{gso}
s_i({\bf x}) = \frac{1}{2\pi L_i^2} \exp \left(-\frac{r^2}{2L_i^2}\right)~~,~~ \frac{R_i(r)}{R} = 1- \exp \left(-\frac{r^2}{2L_i^2}\right)\, ,
\eqx
or, in momentum space, to:
\eq
s_i({\bf k}) =  \exp \left(-\frac{k^2 L_i^2}{2}\right)\, .
\eqx

\section{Numerical solutions and comparison with $x$-space results}

There are two ways to solve the non-linear system  (\ref{mseqs}).
One may use an iterative(recursive) procedure
suggested by the form of the equations, or  treat them
(after discretization) as an algebraic system of polynomial equations
of third order. The two approaches are to some extent complementary.
The recursion turns out to be convergent only in the dispersive phase,
and can therefore be used to determine the inter-phase boundary in parameter
space. The algebraic approach allows to explore also the BH phase
by  generating genuine complex solutions of the system. It can also be cross
checked, of course,  with the recursive method in the dispersive phase.

Both approaches rely on a momentum discretization procedure.  As a first step,
 we introduce new variables which span unit intervals and are thus convenient
for that purpose:
\eqn
x=\frac{1}{1+k_1 L},\;\;\;y=\frac{1}{1+k_2 L},\;\;\;v=\frac{1}{1+k L} .
\eqnx
Here $L$ is the size of the two identical sources (A, B from the previous
Section) or of one of them ($L_1$ in the case C).
In the following we set $L=1$.
In the new variables equations (\ref{mseqs}) read:
\eqn
h(v) &=&  \frac{1}{\pi^2}  \int_T \frac{d x d y}{x(1-x) y(1-y)}\frac{v^2}{(1-v)^2}
\sqrt{\lambda(x,y,v)}  \beta_1(x) \beta_2(y) \, , \nonumber \\
\beta_i(v) &=& s_i(v) + \frac{R^2}{2}  \int_T
\frac{dx dy}{x(1-x) y(1-y)} \sqrt{\lambda(x,y,v)} h(x) \beta_i(y)\, .
 \label{veqs}
 \eqnx
At fixed $v$ the $x,y$ integrals are over the triangular region $T$ which is bounded by
three hyperbolas
\eqn
   0 < v < 1,\;\;\; & 0 < x < 1,\;\;\; & y_{min}(x,v) < y < y_{max}(x,v), \\
  y_{min}(x,v)= \frac{x v}{x + v - x v}&&
  y_{max}(x,v)= \left\{ \begin{array}{cc}
                       \frac{x v}{v - x + x v} & x < v \\
                       \frac{x v}{x - v + x v} & x \ge v
                         \end{array} \right.
\eqnx
Next, we discretize the variables
\eqn
u \longrightarrow u_i = \frac{1}{2 n} + \frac{i-1}{n},\;\;\; i=1,\dots,n,\;\; u=x,y,v \, ,
\eqnx
and turn the integral equations into a set of $3 n$ algebraic equations
\eqn \label{peqs}
f_i & = & s_i +\frac{R^2}{2} \Sigma_{j,k=1}^n w^{(b)}_{i,j,k} f_{2 n +i}f_i  \, ,\\
f_{n+i} & = & s_{n+i} + \frac{R^2}{2} \Sigma_{j,k=1}^n w^{(b)}_{i,j,k} f_{2 n +i}f_{n+i}\, , \\
f_{2n+i} & = &  \frac{1}{\pi^2} \Sigma_{j,k=1}^n w^{(h)}_{i,j,k} f_{i} f_{n+i} \, ,
\eqnx
where the weights $w$ are the discretized versions of the $kernels*measure$
 in the corresponding continuous equations
\eqn
w^{(b)}(i,j,k) &=& \frac{\lambda(x_j,y_k,v_i)}{x_j(1-x_j)y_k(1-y_k)} Area(i,j,k) \, ,\\
w^{(h)}(i,j,k) &=& \frac{\lambda(x_j,y_k,v_i)}{x_j(1-x_j)y_k(1-y_k)(1/v_i -1)^2} Area(i,j,k) \, ,\\
\eqnx
and the $Area(i,j,k)$ is the area of the intersection of a small, $1/n^2$, square  with
the "triangle" $T$ if the center of the square $(x_j,y_k)$ is inside $T$,
and zero if the center is outside $T$.

As already remarked, these equations can be used either to set up
a recursive procedure, or can be directly solved numerically.
The latter approach, dubbed  the ``algebraic method",
works in both the BH and the dispersion phase.

In the first two (A and B) rows of Table \ref{xvsp} we compare a sample of
critical values of $R$ (in units of $L$) as determined by  configuration
\cite{VW08} and momentum space methods.
In the latter $R_c$ is defined as a point where the recursion (\ref{peqs})
diverges. As such it depends on the length of a
trajectory $n$ and in principle requires extrapolation to $n=\infty$.
In the Table we used $ n \le 80 $.

\begin{table}[h]
\begin{center}
\begin{tabular}{cccccccc}
 \hline\hline
  $d$        & & $   0.5   $ & $  1.0  $ &  $ 1.6   $ & $2.5$ & $4.0$ &       \\
   \hline
  A-x      & & $  0.419  $ & $ 0.471 $ &  $ 0.502 $ & $0.528$ & $0.550$  &  \\
  A-p      & & $  0.429  $ & $ 0.476 $ &  $ 0.499 $ & $0.501$ & $0.477$  &  \\
  \hline\hline
  $\sigma$   & & $0.01 $ & $ 0.1 $   & $   0.2   $ & $  0.3  $ &  $ 3.0   $ & \\
  \hline
 B-x  & & $0.615$ & $ 0.572 $ & $  0.525  $ & $ 0.486 $ &  $ 0.470 $  &\\
 B-p  & & $0.058$ & $ 0.436 $ & $  0.501  $ & $ 0.489 $ &  $ 0.476 $  & \\
   \hline\hline
$\rho$ & $0.25$ & $ 0.333 $ & $ 0.5 $ & $ 1.0 $ &  $2.0$  &  $3.0$ &   4.0    \\
   \hline
 C-x  & $.810$ & $ .816 $ & $.821$ & $.823$ & $.821$ &  $.816$  & $.810$\\
 C-p  & $.823$ & $ .833 $ & $.850$ & $.841$ & $.838$ &  $.840$  & $.832$\\
   \hline\hline
\end{tabular}
\end{center}
\caption{$(R/L)_c$ for a range of sizes of the power-like and Gaussian sources:  a comparison between  configuration
 and momentum-space results. A, B and C label sources as discussed in Sect.3.
 In the case C: $\rho=L_2/L_1$ and the critical value of the ratio $2R/(L_1+L_2)$ is shown.}
\label{xvsp}
\end{table}

For extended sources both approaches are consistent meaning that at $n\sim 50 - 80$ momentum space
estimates have already converged. However, for narrower sources, momentum
method requires a yet finer discretization. This is to be expected, since a finite
mesh in momentum, say $\Delta p$, limits the spatial resolution to
$\Delta x > 1/\Delta p$. In such cases one has to extrapolate
numerical data from the case of extended sources as done in \cite{MO}.
In the following, we shall be discussing only homogenous beams where
a finite  $n \sim 50 $ is adequate. Notice again the two special
cases mentioned earlier, namely $d=1 (\infty)$ and $\sigma=\infty (0)$
which correspond to the scattering of homogeneous beams and that of a particle
and a ring. The critical radii for these cases ($R_c \sim .47$ and
$R_c=2^{1/2}3^{-3/4}\sim .62 $ respectively) were obtained in \cite{ACV07} and agree
with the ones quoted in the Table except for the momentum study of the
infinitely narrow sources which was to be expected.

The third row of the Table summarizes the head-on collision of the two
central, gaussian sources with different widths (C).
The problem is symmetric with respect to the interchange
$ L_1 \leftrightarrow L_2 $ therefore we display $R_c$ in
units of $(L_1+L_2)/2$. Agreement between x- and p-space methods
is quite satisfactory. The configuration space technique used in \cite{VW08}
was manifestly symmetric under the interchange of $L_1$ and $L_2$ as reflected
in the Table. However, in the momentum space calculations we have
deliberately used only one source size as a scale. The resulting small
asymmetry gives an idea of  the sensitivity to the discretization parameter $n$
(which was not so large here, $ n \le 20 $).

Finally, we emphasize a very weak dependence of
$ [2R/(L_1+L_2)]_c $ on $\rho$. This confirms the observation, made already in
\cite{VW08}, that the critical line is remarkably linear in the $(L_1,L_2)$
 plane indicating that a simple sum $L_1+L_2$ controls the concentration
 of energy in a large part of parameter space.

\section{Spectrum of emitted gravitons}

\begin{figure}[h]
\epsfig{width=12cm,file=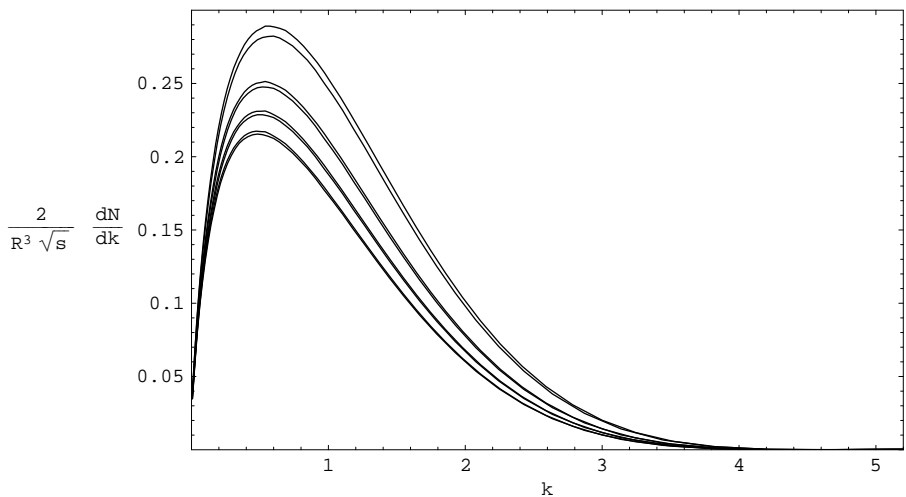}
\vskip-4mm \caption{Scattering of two homogeneous beams of size $L_i =1$. Inclusive spectra $|\kt| |h(\kt)|^2$ of emitted gravitons, as a function of $|\kt|$
close to criticality.
 Bottom to top: $R=0.44, 0.45, 0.46, 0.47$, with $R_c=0.470673$. The two curves for each $R$ are for $n=60$ and $n=70$. }
  \label{dNodk}
\end{figure}

Let us recall, from \cite{ACV07}, that the graviton spectrum is determined in terms of $h(k)$ through:

\eqn\label{dNd3}
\frac{1}{\sigma_T} \frac{d \sigma}{d\kt^2~ dy} \sim  Gs R^2   |h(\kt)|^2 \, .
\eqnx

In Fig.\ref{dNodk} we show a suitably normalized transverse-energy distribution of gravitons, $ |\kt| h(\kt)^2 \sim \frac{d\sigma}{d|\kt|} $, as generated
from the iterations (\ref{peqs}). The two adjacent curves for each $R$ give an idea about the residual
dependence on the ``volume"  $n$. As usual, the convergence with $n$ is slower in the vicinity of the critical point, but the Figure suggests that the behaviour near $R_c$ is rather regular. In particular there is no indication for a buildup of any divergence in the spectrum  as $R \rightarrow R_{c}^-$.

\begin{figure}[h]
\epsfig{width=12cm,file=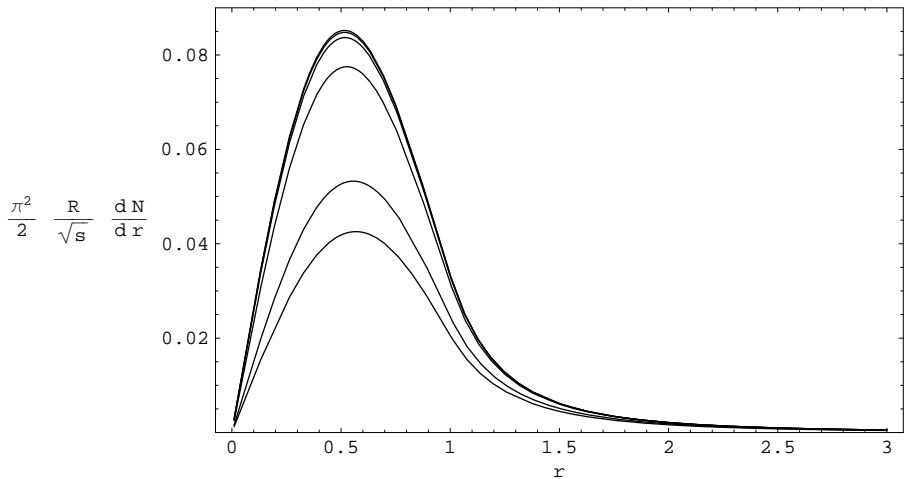}
\vskip-4mm \caption{Same as in the previous figure but in configuration space. The various curves (brom bottom up) correspond to
$R=0.45, 0.46, 0.47, 0.4706, 0.47064, 0.47065$, and $R_c=0.470673$. }
  \label{dNodr}
\end{figure}

A similar conclusion follows from  Fig.\ref{dNodr} where the density profile of the gravitational field
 $(\pi R)^4 r h(r)^2 = r(1-\dot{\rho}(r))^2$ in  transverse distance $r$
is shown. The numerical calculations were done in  configuration
space following \cite{VW08}. There is no ``finite volume", $n$, in this case  and we can
concentrate on the dependence on $R$. Again, no singularity in $r$
develops as $R \rightarrow R_c$ and we can define a smooth
limiting distribution {\em at} the critical point $R_{c}^-$. Note
however, that the dependence on $R$ close to $R_c$ is rather strong
(the  change of $R$ is tiny  for the three uppermost curves),
suggesting that the limiting distribution is attained with a large,
possibly infinite, derivative.

\begin{figure}[h]
\begin{center}
\epsfig{width=12cm,file=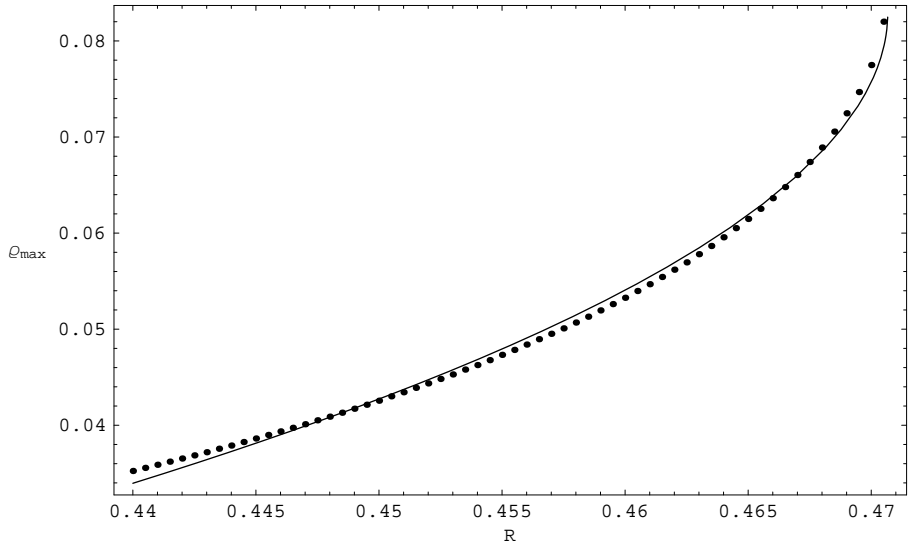}
\end{center}
\vskip-4mm \caption{The maximum of the spatial density of gravitons as a function of $R$. Points are
from solving Eqs.(\ref{aseqs}). The linear fit, $0.083-0.28(R_c-R)^{1/2}$, is also shown.}
  \label{rhomax}
\end{figure}

Figure \ref{rhomax} illustrates yet better  this point. It turns out that
the maximum value of the above density depends on $(R_c-R)$ as a simple
square root. When left unconstrained, the best fitted power was always
within $\pm 1 \%$ of $1/2$, remaining very stable against adding or removing
initial/final data points. The solid line in Fig. \ref{rhomax} shows the fit where
the power was actually fixed to $1/2$. Similarly, we have found that
the width of the distribution also behaves as $ c_1 + c_2 (R_c-R)^{1/2} $,
with finite coefficients $c_1, c_2$.
Therefore, indeed, the limiting distribution exists and is approached
with an infinite derivative w.r.t.  $R$. At the transition point, gravitons are emitted
preferentially  from half the  distance from  the source's  edges.

In \cite{VW08} we have already analyzed the total multiplicity of emitted gravitons since it is related to the  derivative of the action
 with respect to $R$. We found that it approaches a finite constant at $R_{c}^-$ with a square root
branch point at $R_c$ (see figure 5 of \cite{VW08}), with a ``best fit" given by $0.138-0.46(R_c-R)^{0.523}$.
This is hardly surprising given the above results for the differential distribution.

To close the circle we have also
compared the Fourier transform of the $p$-space solution $h(k)$ with
$h(r)$ obtained directly from the solution $\rho(r)$ in
$x$-space. The two agree locally within  2-3 \%, the discrepancy being caused
again by the finite discretization in momentum space. This is
reassuring, not only in confirming again the consistency of the
whole procedure, but also because, due to the infrared behaviour,
transition between $x$- and $p$- representations is rather subtle.
In particular,
we find ({\em cf.} again Fig. \ref{dNodk} )that  the emission amplitude $h(k)$ (and also $\beta_i(k)$)
diverges at small momenta as
\eqn
h(k) \sim \frac{1}{\sqrt{k}}\, ,
\eqnx
and is exponentially damped at large $k$. Consequently,  the action (\ref{pspA})
is IR divergent and, even if the divergence is physically irrelevant
({\em cf.} the infinite Coulomb phase), has to be treated with care numerically.

\begin{figure}[h]
\begin{center}
\epsfig{width=12cm,file=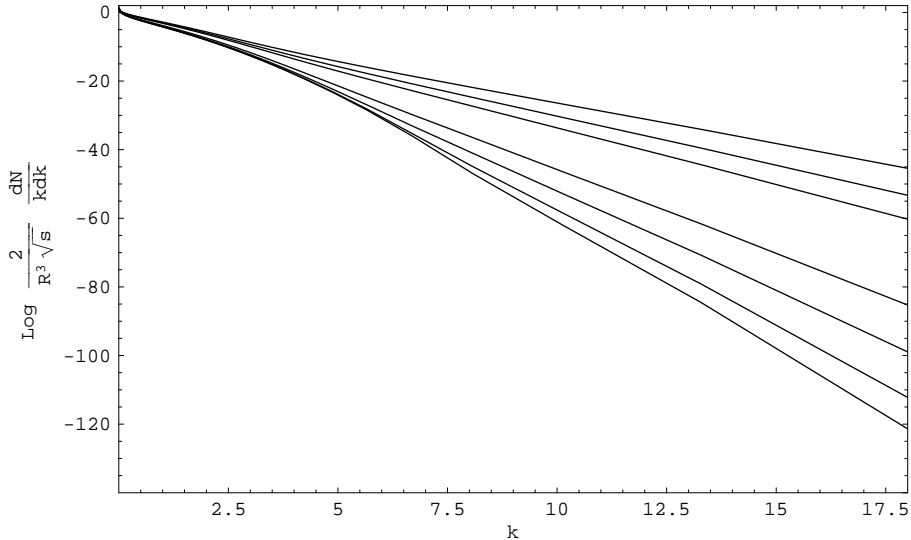}
\end{center}
\vskip-4mm \caption{A logarithmic  plot of the distribution (\ref{dNd3})
for  $R=.10, .20,.40,60,.80,.82,.83 $ (bottom to top).}
  \label{dNforRs}
\end{figure}

\begin{figure}[h]
\begin{center}
\epsfig{width=10cm,file=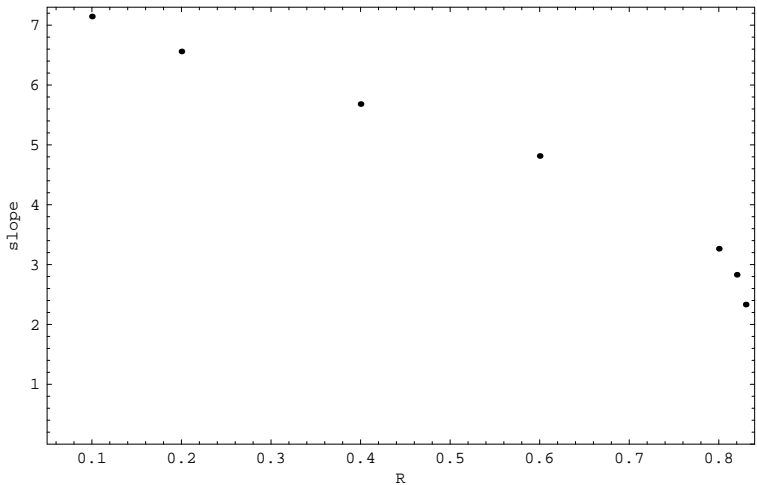}
\end{center}
\vskip-4mm \caption{R dependence of the slopes from
 the previous figure (determined locally from $10 < k < 12.5$).}
  \label{slopesofR}
\end{figure}

The large $k$ behaviour of the spectrum can be qualitatively assessed from
Fig. \ref{dNforRs}. The distribution resembles much more an exponential than
the gaussian shape of the sources we have put in (this is the C case of Sect.3).
Actually, a rather interesting structure emerges, as shown in that Figure.
Let us use units in which $L=1$ and  increase $R$ (i.e. the energy),
starting from very small values. In that perturbative region  the spectrum
appears to consist of {\it two} exponentials separated by a  ``knee",
i.e. the slope  at small $k$, is smaller than the one at large $k$.
However, as we increase $R$ towards its critical value, the knee tends to
disappear leaving behind an almost perfect exponential $exp(-b|k|)$.
The slope $b$ of the exponential (determined mainly by $L$ in the
perturbative region) now strongly depends on $(R-R_c)$.
Fig.\ref{slopesofR}  illustrates the points we just made and indicates
that the slope also has
a rather singular behaviour, possibly of the form
$b \sim c_1 + c_2 (R-R_c)^{\gamma}$.

One can finally try to determine the total multiplicity by integrating the
spectra (either in $x$ or in $p$ space). At small $R/L$ one finds:
\eqn
\langle N \rangle \sim Gs (R/L)^2 \, .
\eqnx
Given that the average transverse energy is $O(1/L)$ this corresponds to an average total transverse energy in the emitted gravitons given by:
\eqn
\langle E_T \rangle \sim \sqrt{s} (R/L)^3 \ll \sqrt{s} \, .
\eqnx
However, as one approaches $R_c$, this quantity clearly becomes $O(\sqrt{s})$
possibly implying that the transverse energy becomes a good estimate of
the total radiated energy but also, unfortunately, that imposing
energy-conservation, i.e. taking into account the back-reaction on the sources,
becomes mandatory.

\section{Beyond the critical point: a first attempt}

There is obviously  much interest in extending the numerical analysis into a BH phase.
Although  precise solutions for $R>R_c$ are not available at this stage, we can get an approximate picture
of their complicated  structure using the momentum-space approach. As already explained
 in Sect.4,  the discretized
equations of motion (\ref{veqs}) can be solved exactly (though numerically) as an algebraic system. This can be done
for arbitrary $R$, without any use of the iterative  approach. The price is, of course, that we deal with
a non-linear system of $3n$ equations for $3n$ variables, which is quite challenging even numerically.
Restricting to identical sources reduces a problem to $n$ variables.
Still there are $3^n$ solutions, and Mathematica has to generate all of them before we
can choose the physically acceptable ones. To produce all $729$ solutions for $n=6$ takes about $2.5$ hours hence, in practice,
the method is limited at present to $n \le 5$. Still, even such a crude discretization reasonably reproduces
the main features of, say, the momentum distribution.  Therefore we go ahead
 with the simplified problem here and look for complex solutions
above $R_c$ leaving the, obviously possible, refinements for the future.

\begin{figure}[h]
\begin{center}
\epsfig{width=12cm,file=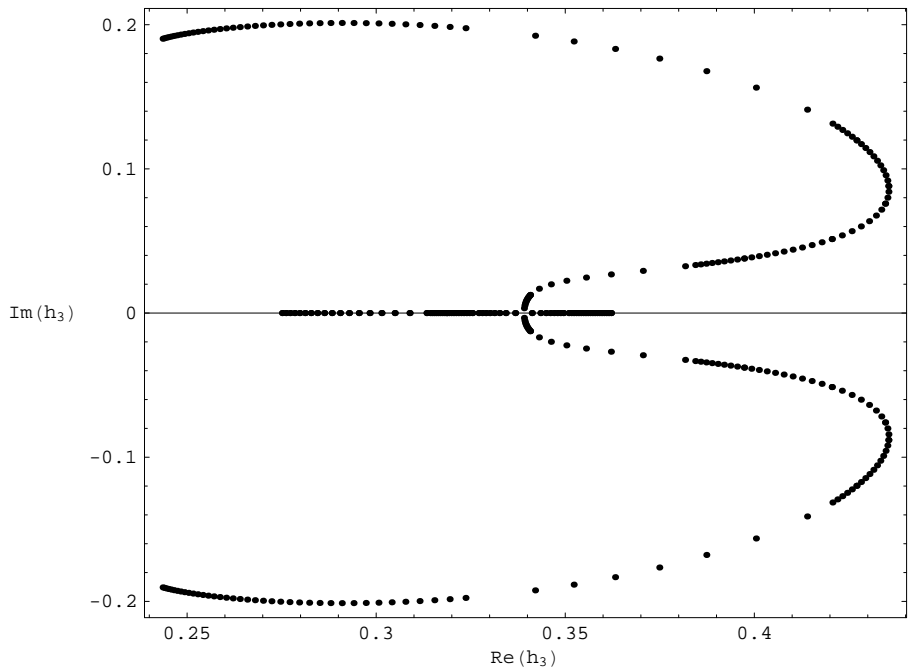}
\end{center}
\vskip-4mm \caption{ Two complex trajectories $h(k)$ (at $k=k_3=2.$) parametrized by $R$ moving
through the critical point.}
  \label{traj2}
\end{figure}

Figure \ref{traj2} shows the behaviour of two complex trajectories\footnote{By a trajectory
we mean here a path traced by $h(k)$, at one value of $k$, while changing $R$.} as R moves from the dispersive to the BH phase.
We show only a discrete series of points with varying step $\Delta R$ (see below) to give also an idea
of the speed (in $R$) along a trajectory. For $n=5$, the critical point is at $R_c= 0.3193$.
We begin by following the left trajectory (also reproduced by the recursion in this region)
deep inside  the dispersive phase (at $R=0.15$): it moves towards its critical value $x_c$ along the real axis, as expected.
Checking the stability matrix at the fixed point we verified that, indeed, this is the {\em only} solution which is stable
against the iterations (\ref{peqs}). The second solution (which starts
above $x_c$ in the Figure)
 is also real, but unstable. We  have chosen it to be the one that matches
 the recursive solution at $x_c$.
The sampling points are separated by $\Delta R =0.01$ at the beginning of the
recursive solution and then,
as we approach $R_c$, the coarse graining was reduced to $\Delta R=0.001$.
 One readily sees that while approaching $x_c$ the variation w.r.t. $R$
increases in agreement with the  findings of Sect. 5.
For the unstable trajectory $\Delta R$ was set to $0.001$ from the beginning
since we started tracing it much closer to $R_c$.
Again, its velocity increases as we go towards the critical point.
As soon as one goes above  $R_c$ both trajectories acquire (complex conjugate) imaginary pats $y(R)$. Closer inspection shows
that these imaginary parts grow like a square root, $\sqrt{x - x_c}$ -- typically for a threshold behaviour. Since both
 trajectories are complex conjugate to each other, Fig. \ref{traj2} implies that the real part
 of the left trajectory continues to increase while that of the right trajectory changes from decreasing to increasing
 as we go through  $R_c$. Interestingly, the trajectories reveal a rich structure even when we go deeper inside  the BH phase.
 We have followed them up to $R=0.65$ changing the R-resolution:
 $\Delta R = 0.001 \rightarrow0 .01 \rightarrow 0.005 \rightarrow 0.01 $ and finally to $\Delta R=0.003$ along the last segments.
In these regions the trajectories appear to saturate. It remains to be seen if all these detailed features,
like bending or saturation, are generic or are just artefacts of our small value of $n$. For example, we know that increasing
$n$ would move $R_c$ by more than 30~\%.

However, one thing is clear: for $R > R_c$  classical solutions of
(\ref{peqs})
(and consequently also the on shell effective action) develop imaginary parts.
This is turn implies a new absorption of the elastic amplitude (on top of the one due to graviton production) calling for the opening of some new channel.
 The whole mechanism is somewhat reminiscent of the classic discussion of
 the ``decay of the false vacuum" by Coleman and de Luccia \cite{Coleman} as a tunnelling process described through the contribution of complex saddle points to the functional integral. Use of similar ideas in this new context is presently under investigation \cite{CC}.

  Let us conclude by stressing that the  non-linearities captured by equations (\ref{peqs}) are essential
  for the
 above instabilities to occur. The gravitational attraction alone exists already in the lowest Born diagram,
 but it is not sufficient to produce the non-linearities of the metric that are essential for the buildup of CTS.
 Our results confirm that, instead, the class of diagrams selected in \cite{ACV07} appear to be  sufficient
for bringing out  the emergence of such phenomena.


\section*{Acknowledgements}
We wish to thank   M. Ciafaloni for interesting discussions and  for communicating to us some preliminary results from ref. \cite{CC} prior to publication.  We also acknowledge the warm hospitality of the Galileo Galilei Institute in Arcetri (Florence, Italy) during the completion phase of this work.


\begin{thebibliography}{99}
\bibitem{VW08} G. Veneziano and J. Wosiek, {\it Exploring an S-matrix for gravitational collapse} (hep-th/0804.3321)
\bibitem{ACV07}
D. Amati, M. Ciafaloni and G. Veneziano, {\em JHEP} {\bf 02} (2008) 049 (hep-th/07121209).
\bibitem{ACV}
D. Amati, M. Ciafaloni and G. Veneziano, {\em Phys. Lett.} {\bf B197} (1987) 81;
{\em Int. J. Mod. Phys. A} {\bf 3} (1988) 1615.
\bibitem{Lip} L. N. Lipatov, {\it Nucl. Phys.} {\bf B365} (1991) 314;\\ R. Kirschner and L. Szymanowski, {\it Phys. Rev.} {\bf D52} (1995) 2333.
\bibitem{ACV93} D. Amati, M. Ciafaloni and G. Veneziano, {\em Nucl. Phys.} {\bf B403} (1993) 707;
 {\em Nucl. Phys.} {\bf B347} (1990) 550.
  \bibitem{MO} G. Marchesini and E. Onofri, {\em High energy gravitational scattering: a numerical study},
hep-th/0803.0250.
\bibitem{EG}
D. M. Eardley and S. B. Giddings, {\em Phys. Rev.} {\bf D66} (2002) 044011; \\
H. Yoshino and Y. Nambu, {\it Phys. Rev.} {\bf D67} (2003) 024009; \\
S. B. Giddings and V. S. Rychkov, {\em Phys. Rev.} {\bf D70} (2004) 104026.
 \bibitem{KV} E. Kohlprath and G. Veneziano,  {\em JHEP} {\bf 0206} (2002) 057.
\bibitem{AE} A. M.  Abrahams and C. R. Evans, {\em Phys. Rev. Lett.} {\bf 70} (1993) 2980;  {\em Phys. Rev.} {\bf D49} (1994) 3998.
\bibitem{numerical} M. W. Choptuik, E. W. Hirschmann, S. L. Liebling and  F. Pretorius, {\it Critical Collapse of the Massless Scalar Field in Axisymmetry}, {\em Phys. Rev.}  {\bf D68} (2003) 044007.
\bibitem{Chop} M. W. Choptuik, {\it Phys. Rev. Lett.} {\bf 70} (1993) 9.
For a  review, see e.g. C. Gundlach, {\it  Critical Phenomena in Gravitational Collapse}, gr-qc/ 0210101.
\bibitem{Coleman} S. Coleman and F. de Luccia, {\em Phys. Rev.} {\bf D21} (1980) 3305.
\bibitem{CC}  M. Ciafaloni and D. Colferai, private communication.
\end{thebibliography}
\end{document}